\documentclass[aps,prd,preprintnumbers,showpacs,showkeys,nofootinbib,
superscriptaddress,fleqn,floatfix,tightenlines,twocolumn]{revtex4-1}
\usepackage{amsmath,amsfonts,amssymb,amscd,amsxtra,amsthm}
\usepackage{graphicx}  
\usepackage{epstopdf}
\usepackage{dcolumn}  
\usepackage{bm}          
\usepackage{slashed}
\usepackage[utf8]{inputenc} 

\usepackage[normalem]{ulem} 
\usepackage[dvipsnames]{xcolor} 
\renewcommand\sout{\bgroup \color{red} \ULdepth=-.5ex \ULset}

\newcommand\T{\rule{0pt}{2.6ex}}       

\begin{document}
\preprint{INHA-NTG-12/2018}
\title{Instanton effects on charmonium states} 

\author{Ulugbek~Yakhshiev}
\email{yakhshiev@inha.ac.kr}
\affiliation{Department of Physics, Inha University,
Incheon 22212, Republic of Korea}

\author{Hyun-Chul~Kim}
\email{hchkim@inha.ac.kr}
\affiliation{Department of Physics, Inha University,
Incheon 22212, Republic of Korea}
\affiliation{Advanced Science Research Center, Japan Atomic Energy
  Agency, Shirakata, Tokai, Ibaraki, 319-1195, Japan} 
\affiliation{School of Physics, Korea Institute for Advanced Study
  (KIAS), Seoul 02455, Republic of Korea} 

\author{Emiko Hiyama}
\email{hiyama@phys.kyushu-u.ac.jp}
\affiliation{Advanced Science Research Center, Japan Atomic Energy
  Agency, Shirakata, Tokai, Ibaraki, 319-1195, Japan} 
\affiliation{Department of Physics,Kyushu
  University,819-0395,Fukuoka,Japan}    
\affiliation{RIKEN Nishina Center, RIKEN, 2-1 Hirosawa, 351-0115
  Saitama, Japan} 

\date{\today}

\begin{abstract}
The instanton effects on the charmonium spectrum is discussed in 
the framework of nonrelativistic potential model.
The results from constituent quark model without inclusion of 
instanton effects is compared with the results for `the potential 
from constituent quark model + the contribution
from the instanton liquid model'. We consider two models with 
the corresponding instanton potentials and discuss their relevance
to explanations of the origin phenomenological parameters 
used in the nonrelativistic potential models.
We also present the universal instanton potential in the parametrized 
form which can be useful in practical calculations.
\end{abstract}


\maketitle
\section{Introduction}
Physics of charmonia has entered a new era, since the finding of the
first narrow exotic charmonium was reported
by the Belle Collaboration~\cite{Choi:2003ue}. Many 
narrow exotic charmonium states have been consecutively
observed~\cite{Aubert:2004ns, Aubert:2005rm, Abe:2007jna, 
  Choi:2007wga, Belle:2011aa, Liu:2013dau, Ablikim:2013mio,
  Ablikim:2013wzq, Aaij:2013zoa,Ablikim:2013xfr,Aaij:2014jqa, 
  Aaij:2015zxa} and were coined collectively the XYZ mesons (see
recent reviews~\cite{Yuan:2015kya,Yuan:2015ztu}). These new exotic
charmonium states have drawn considerable attention (see for example
the reviews in Refs.~\cite{Swanson:2006st, Eichten:2007qx,
  Voloshin:2007dx, Brambilla:2010cs,Olsen:2014qna, Chen:2016spr}) 
and also brought about multifaceted view point on conventional
charmonium states.  

Theoretically, the quantum-mechanical potential models provide an  
easy but very effective way of describing the charmonum
spectrum~\cite{Eichten:1974af, Eichten:1978tg, Eichten:2007qx,
  Voloshin:2007dx, Brambilla:2010cs}. In a standard approach there are
basically two main contributions to the heavy-quark potential for the
charmonium system: the Coulomb-like potential and the phenomenological
quark-confining one. The Coulomb-like potential originates from
one-gluon exchange (OGE) between a heavy quark ($Q$) and a heavy
anti-quark ($\bar{Q}$)~\cite{Susskind:1976pi, Appelquist:1977tw, 
  Appelquist:1977es, Fischler:1977yf}, based on 
perturbative quantum chromodynamics (pQCD). Note that the static
Coulmob-like potential was scrutinized already to 
higher-order corrections from pQCD~\cite{Peter:1996ig, 
  Peter:1997me, Schroder:1998vy, Smirnov:2009fh,
  Anzai:2009tm}. By nature of pQCD, the Coulomb-like interactions are 
supposed to govern the short-range physics of the charmonia. 
At large distances the strength of the Coulomb-like interaction 
decreases. However, the presence of the quark-confining potential
makes the strength of the total interaction increase. This is due to
the fact that any quark inside a charmonium is ordained to be confined
in it, effects of the quark confinement~\cite{Wilson:1974sk} are
necessarily involved. The heavy-quark potential for the quark
confinement can be obtained at least phenomenologically from the
Wilson loop, which rises linearly at large
distances~\cite{Eichten:1974af, Eichten:1978tg}.   On the
other hand, as the quark and the anti-quark start to recede from each
other the certain nonperturbative contributions come into play.

Recently, we examined yet another nonperturbative effects on the
mass spectrum of the charmonia from the instanton vacuum of
QCD~\cite{Turimov:2016adx}.  The central part of the heavy-quark
potential was already derived by Diakonov et
al.~\cite{Diakonov:1989un}, based on the instanton liquid  
model for the QCD vacuum~\cite{Diakonov:1983hh, Diakonov:1985eg,
  Diakonov:2002fq}.  The spin-dependent part of the instanton-induced
potential can be easily obtained by employing the method of Eichten
and Feinberg~\cite{Eichten:1980mw}. There are two intrinsic
parameters of characterizing the instanton vacuum, i.e. the average
size of the instanton $\rho$ and the average distance
$R$ between instantons.  Their numerical values were
estimated to be $\rho\approx 0.33\,\mathrm{fm}$ and
$R\approx 1 \,\mathrm{fm}$~\cite{Shuryak:1981ff,
  Diakonov:1983hh, Diakonov:1985eg}. However, these values are just
the approximated ones. For example, Refs.~\cite{Kim:2005jc,
  Goeke:2007bj,Goeke:2007nc} considered $1/N_c$  meson-loop
contributions in the light-quark sector and found it necessarily to
readjust the values of parameters as
$\rho\simeq0.35\,\mathrm{fm}$ and $R\simeq0.86\,\mathrm{fm}$.   In
Ref.~\cite{Turimov:2016adx}, we scrutinized the dependence of the
heavy-quark potential from the instanton vacuum. In the present work,
we combine the instanton-induced heavy-quark potential with the
Coulomb-like and quark-confinement potentials and investigate
explicitly the instanton effects on the mass spectrum of the
charmonia.  

The paper is organized as follows: In 
Section~\ref{sec:VQQbar}, we explain briefly the nonrelativistic
heavy-quark potential model consisting of the color Coulomb-like
potential and the linear scalar potential for quark
confinement~\cite{Barnes:2005pb} and discuss the contribution to the
total heavy-quark potential from the instanton vacuum. In 
Section~\ref{sec:NumMeth}, the Gaussian expansion method for a
numerical calculation will shortly be described.  
In Section~\ref{sec:Results} we present the results and discuss them. 
The final Section~\ref{sect:Summary} is devoted to the summary and
conclusions of this work.
\section{Heavy-quark potential}
\label{sec:VQQbar}

The standard heavy-quark potential consists of the four different 
terms, written as 
\begin{align}
\label{VQQbar}
V_{Q\bar{Q}}(\bm{r})& = V_C(r)
+V_{SS}(r)
(\bm{S}_Q\!\cdot\!\bm{S}_{\bar Q})\cr
&+V_{LS}(r)(\bm{L}\cdot\bm{S}) \\
&+V_{T}(r)\left[
3(\bm{S}_Q\!\cdot\!\bm{n})(\bm{S}_{\bar Q}\!\cdot\! \bm{n})-\bm{S}_Q 
    \cdot\bm{S}_{\bar Q}\right],\nonumber 
\end{align}
where $V_C$, $V_{SS}$, $V_{LS}$ and $V_T$ represent respectively the 
central, spin-spin, spin-orbit and tensor parts of the heavy-quark
potential. $\bm{S}_Q$, $\bm{S}_{\bar{Q}}$, and $\bm{L}$ denote the
spin operator of a heavy quark, that of a heavy anti-quark, and the
relative orbital angular momentum operator, respectively. The 
unit vector $\bm{n}$ in three dimensional ordinary space is chosen in
direction of the line joining the centers of the heavy quark and heavy
anti-quark. All contributions to the total $Q\bar{Q}$  
potential can be constructed within the framework of nonrelativistic
potential approaches. As mentioned previously, the color Coulomb-like   
vector potential arises from OGE between $Q$ and $\bar{Q}$ while the
linear scalar potential is constructed phenomenologically from the
area law of the Wilson loop~\cite{Wilson:1974sk} for the quark
confinement. We denote the corresponding \textit{OGE vector $+$ scalar
confining potential} as $V^{\mathrm{(P)}}$. In addition, 
we introduce the nonperturbative potential $V^{\mathrm{(NP)}}$ derived
from the instanton vacuum~\cite{Turimov:2016adx}. Consequently, the
heavy-quark potential in the present work is written as  
\begin{align}
V(r)=V^{\mathrm{(P)}}(r)+V^{\mathrm{(NP)}}(r).
\label{eq:2}
\end{align}
We will compute the mass spectrum of the charmonia based on
the potential in 
Eq.~\eqref{eq:2}. However, we want to emphasize that we will not 
carry out a fine tuning to reproduce the experimental data, since it
is of greater importance to understand the nonperturbative effects 
coming from the instanton-induced potential and physical implications
of the parameters involved. 
\subsection{Heavy-quark potential}
A nonrelativistic potential model~\cite{Barnes:2005pb} provides a
minimal theoretical framework to describe the charmonia.  
The central part of the heavy-quark potential is expressed as 
\begin{align}
\label{eq:phenVC}
V_C^{\mathrm{(P)}}(r)=\kappa r-\frac{4\alpha_s}{3r}, 
\end{align}
where the first term expresses the linear scalar potential for the
quark confinement and the second one comes from OGE, respectively.   
Here 
$\kappa$ stands for the parameter of the string tension, of which the
numerical value can be approximately determined by reproducing the
mass spectrum of the charmonia. 
The parameter $\alpha_s$ denotes the running strong coupling
constant of pQCD, of which the value is theoretically well-known. 
We fix the scale of $\alpha_s$ at the mass of the charm quark.
We will discuss physical implications
of these two parameters later. 

The spin-dependent parts can be obtained from the central
potential~\eqref{eq:phenVC}
\begin{align}
\label{eq:phVss}
V_{SS}^{\rm (P)}(r) 
&=\frac{32\pi\alpha_s}{9m_Q^2} \tilde{\delta}_\sigma(r),\\
V_{LS}^{\rm (P)}(r)&=
\frac{1}{2m_Q^2}\left(\frac{4\alpha_s}{r^3}-\frac{\kappa}{r}\right),
\label{eq:VLSP}\\
V_{T}^{\rm (P)}(r)&=\frac{4\alpha_s}{m_Q^2r^3},
\label{eq:VTP}
\end{align}
which appear from the next-to-leading order in the expansion of the
heavy quark mass $m_Q$. So, the spin-dependent potentials are
proportional to $1/m_Q^2$, respectively. 
In the present work, the value of the charm-quark mass $m_c$ will be 
determined by including the instanton effects. 
A \textit{smeared} Dirac delta function $\tilde{\delta}_\sigma$ in 
Eq.~\eqref{eq:phVss} is written in the Gaussian form
\begin{align}
\tilde{\delta}_\sigma(r)=\left(\frac{\sigma}{\sqrt\pi}\right)^3
e^{-\sigma^2r^2},
\end{align}
where $\sigma$ is the smearing parameter that can be determined
phenomenologically.
\subsection{Instanton-induced potential}
In addition to the potential given in Eq.~\eqref{eq:phenVC}, we
introduce the instanton-induced potential, which was already derived
in Refs.~\cite{Diakonov:1989un, Turimov:2016adx}. The explicit form
of the potential is expressed as 
\begin{align}
\label{eq:LO_pot}
V_C^{\mathrm{(NP)}}(r) =
\frac{4\pi\rho^{\,3}}{N_cR^4}\,\mathcal{I}\left(\frac{r}{\rho}\right),
\end{align}
where $N_c$ denotes the number of colors, $\rho$ and $R$ stand for the 
average instanton size and average inter-instanton distance.
The dimensionless integral $\mathcal{I}(x)$ is given as a function of
the dimensionless variable $x$ 
\begin{align}
\label{eq:dimint}
\mathcal{I}(x)&=\int_0^\infty y^2dy\int_{-1}^1dt\,\left\{1-\cos\left(\frac{\pi
        y}{\sqrt{y^2+1}}\right)\right.\cr 
&\times
\cos\left(\pi\sqrt{\frac{y^2+x^2+2xyt}{y^2+x^2+2xyt+1}}\,\right)\cr
&-\frac{y+xt}{\sqrt{y^2+x^2+2xyt}}\sin\left(\!\frac{\pi
  y}{\sqrt{y^2+1}}\right)\cr 
&\times\left.
\sin\left(\pi\sqrt{\frac{y^2+x^2+2xyt}{y^2+x^2+2xyt+1}}\right)
\right\}.
\end{align}
Though it is possible to compute Eq.~\eqref{eq:dimint} numerically 
given $x$, it is more convenient to parametrize the integral
$\mathcal{I}(x)$ such that one can easily make it useful for a practical
calculation. We obtain a suitable parametrization as follows
\begin{align}
\label{eq:Ixpar}
\tilde{\mathcal{I}}(x) &=
\mathcal{I}_0\Big[1+\sum_{i=1}^3a_ix^{2(i-1)}e^{-b_i x^2}\cr 
&+\frac{a_4}{x}\left(1-e^{-b_4 x^3}\right)
\Big],
\end{align}
where the prefactor $\mathcal{I}_0$ is expressed in terms of the
Bessel functions 
\begin{align}
\mathcal{I}_0 = -\frac{2 \pi^3 }{3}
  \left(J_0(\pi)+\frac1{\pi}J_1(\pi)\right) 
\end{align}
and the parameters $a_i$'s and $b_i$'s are summarized 
in the following matrix forms
\begin{align}
a=\left(
\begin{array}{l}
-1\\ \quad\! 0.10184\\
\quad\! 0.00064\\ -1.11267
\end{array}\right),
\quad b=
\left(
\begin{array}{l}
0.25135\\
0.70255\\
0.18625\\
0.04644
\end{array}\right).
\end{align}

One can inspect whether the parametrized function $\tilde{\mathcal{I}}$  
in Eq.\,\eqref{eq:Ixpar} yields the correct limiting values of the
original integral $\mathcal{I}(x)$ in Eq.\,\eqref{eq:dimint} at
$x=0$ and $x\to\infty$. That is, the following relations  
\begin{align}
\lim_{x\rightarrow 0}\mathcal{I}(x)& = \lim_{x\rightarrow
  0}\tilde{\mathcal{I}}(x) = 0,\cr
\lim_{x\rightarrow \infty} \mathcal{I}(x) &= \lim_{x\rightarrow
  \infty}\tilde{\mathcal{I}}(x) = \mathcal{I}_0.\nonumber 
\end{align} 
are well satisfied. At small $x\ll 1$, $\mathcal{I}(x)$ can be analytically
evaluated as 
\begin{align}
\mathcal{I} (x) & \simeq \frac{\pi^2}{3}
                  \left[\frac{\pi}{16}-J_1(2\pi)\right]x^{2}\cr
&-\pi\left[\frac{\pi^2(438+7\pi^2)}{30720} +\frac{J_2(2\pi)}{80}\right]x^4\cr
&=1.34467x^2-0.500508x^4,
\label{Ix0}
\end{align}
while the parametrization of Eq.~\eqref{eq:Ixpar} gives 
the result
\begin{align}
\tilde{\mathcal{I}} (x)&\simeq \mathcal{I}_0(-a_1b_1+a_2+a_4b_4)x^2\cr   
&+\frac{\mathcal{I}_0}{2}(a_1b_1^2-2a_2b_2+2a_3)x^4\cr
&=1.3316 x^2 - 0.452657 x^4.
\end{align}
At large $x$, the limiting results of $\mathcal{I}(x)$ and
$\tilde{\mathcal{I}}(x)$ are produced as 
\begin{align}
\mathcal{I} (x) &\simeq \mathcal{I}_0 -  \frac{\pi^2}{2x}
=4.41625 - \frac{4.9348}x,\\
\tilde{\mathcal{I}}(x)& \simeq \mathcal{I}_0\left(1
                       +\frac{a_4}{x}\right)
                       =4.41625 - \frac{4.91384}x,   
                       \label{Ipxinf}
\end{align}
respectively. From Eqs.\,\eqref{Ix0}--\eqref{Ipxinf}, one can see that
the corresponding coefficients of the asymptotic forms are very close
to each other.  

\begin{figure}[htp]
\begin{centering}
\includegraphics[scale=0.85]{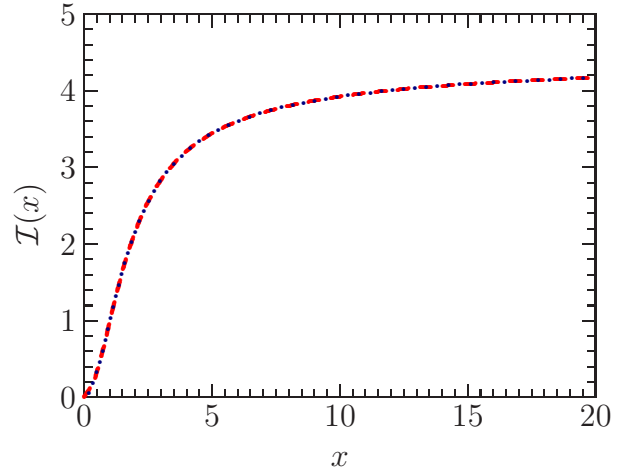}
\end{centering}
\caption{(Color online)
The dimensionless integral $\mathcal{I}(x)$ from the instanton
vacuum. The numerical result of Eq.\,\eqref{eq:dimint} is depicted as 
the red dashed curve whereas that of parametrization given in
Eq.\,\eqref{eq:Ixpar} is drawn as the blue dotted curve.}
\label{fig:1}
\end{figure}

In general, the parametrization in Eq.\,\eqref{eq:Ixpar}
interpolates the numerical value of the integral $\mathcal{I}(x)$  
with a very good accuracy in the whole range of $x$. 
In Fig.~\ref{fig:1}, the numerical result of Eq.\,\eqref{eq:dimint}
is compared with that from the parametrization in
Eq.\,\eqref{eq:Ixpar}.  As shown in Fig.~\ref{fig:1},
they overlap completely each other in the whole range of $x$.  
The maximal value of a relative error is located only at small $x$
region and does not exceed the value  
\begin{align}
\lim_{x\rightarrow 0}&
\frac{\mathcal{I} (x) - \tilde{\mathcal{I}} (x) }{\mathcal{I} (x)}=1
                       \cr 
&-I_0(b_1+a_2+a_4b_4)\left\{\frac{\pi^2}{3} \left[\frac{\pi}{16} -
  J_1(2\pi)\right]\right\}^{-1} \cr 
&= 0.00972.
\end{align}
At large values of $x\rightarrow \infty$, the relative error decreases
according to the formula
\begin{align}
\frac{\mathcal{I}(x)-\tilde{\mathcal{I}}(x)}{\mathcal{I}(x)} & \approx 
-\frac{0.00475}{x-1.11742}.
\end{align}

The spin-dependent parts of the heavy-quark potential from the
instanton vacuum can be obtained from the central part 
according to the following formulas
\begin{align}
\label{eq:VSS}
V^{\rm (NP)}_{SS}(r)&=\frac{1}{3m_Q^2}\nabla^2 
V^{\rm (NP)}_C(r),\\
\label{eq:VLS}
V^{\rm (NP)}_{LS}(r)&=\frac{1}{2m_Q^2}\frac1{r}
\frac{dV^{\rm (NP)}_C(r)}{dr},
\\
\label{eq:VT}
V^{\rm (NP)}_{T}(r)&=\frac{1}{3m_Q^2}
\left(\frac{1}{r}\frac{dV^{\rm (NP)}_C(r)}{dr}
-\frac{d^2V^{\rm (NP)}_C(r)}{dr^2}\right).
\end{align}
Using the parametrization we introduced above, 
we can compute almost all integrations for the instanton-induced
potential analytically. The merit of the parametrization given in 
Eq.\,\eqref{eq:Ixpar} is not limited to the simple calculation of the
integration $\mathcal{I}(x)$. In fact, the instanton-induced central
potential is expressed in terms of $\mathcal{I}(x)$ with all other
instanton parameters factored out. It means that $\mathcal{I}(x)$  
can be used in the universal way for any set of the instanton
parameters.  So, the spin-dependent potentials in
Eqs.\,\eqref{eq:VSS}--\eqref{eq:VT} can be also expressed in terms of
$\mathcal{I}(x)$ for any heavy-quark degrees of freedom, i.e. for the
charmonia or bottomonia.   

The instanton liquid model for the QCD vacuum has two important
intrinsic parameters: the average size of the instanton
$\rho$ and the average inter-distance $R$ between
instantons. In fact, the $Q\bar{Q}$ potential is sensitive to them.
Thus, we will consider three different sets of instanton parameters by
changing them in a permissible manner. 
Model~I (M-I) uses the original values of the instanton parameters:  
$\rho \simeq 0.33\,\mathrm{fm}$ and 
$R \simeq 1\,\mathrm{fm}$. 
However, $\rho$ and $R$ can be changed in a different situation. 
For example, Refs.~\cite{Kim:2005jc, Goeke:2007bj,Goeke:2007nc}
considered the $1/N_c$ meson-loop contributions in the light-quark
sector and found it necessary to readjust them as
$\rho\simeq0.35\,\mathrm{fm}$ and $R\simeq0.856\,\mathrm{fm}$.  
Model IIa (M-IIa) employs these values as in
Ref.~\cite{Turimov:2016adx}. In lattice QCD, the instanton vacuum was
simulated and the following values were suggested: $\rho \approx
0.36\,\mathrm{fm}$ and $R \approx 0.89
\,\mathrm{fm}$~\cite{Chu:1994vi, Negele:1998ev, DeGrand:2001tm,
  Faccioli:2003qz}, which are almost the same as those with the
$1/N_c$ meson-loop corrections. Therefore, the model with this set is
referred as Model~IIb (M-IIb) also as in Ref.\,\cite{Turimov:2016adx}. 
The parameter dependence of the potential can be easily understood
from the form of the leading-order potential expressed in
Eq.\,(\ref{eq:LO_pot}). 
While the prefactor $\rho^3/R^4N_c$, which includes both 
the parameters, governs the overall strength of the
potential, its range is dictated only by the average instanton size
$\rho$ through the dimensionless integral $\mathcal{I}(r/\rho)$.  

\section{Numerical method}
\label{sec:NumMeth} 
In order to evaluate the bound states in the spectrum of the
quarkonia, we have to solve the Schr\"{o}dinger
equation~\cite{Quigg:1979vr} 
\begin{align}
(\hat H-E)|\Psi_{JJ_3}\rangle=0, 
\label{eq:Schroe}
\end{align}
where $\hat H$ is the Hamiltonian operator and $|\Psi_{JJ_3}\rangle$
represents the corresponding state vector with the total angular
momentum $J$ and its third component $J_3$.  
The projection of the state vector 
$\langle \bm{r}|\Psi_{JJ_3}\rangle$ will reproduce the representation
of the Hamiltonian in coordinate space  
\begin{align}
\hat H(\bm{r})=-\frac{\hbar^2}{m_Q}\nabla^2+V_{Q\bar
  Q}(\bm{r}), 
\end{align}
where $\mu_Q$ arises from the reduced mass of the quarkonium 
system. The matrix elements of the $Q\bar{Q}$ potential in the 
standard basis $ | {}^{2S+1}L_J\rangle$,
which is given in terms of the total spin $S$, the orbital angular
momentum $L$, and the total angular momentum $J$ satisfying 
the relation $\bm{J}= \bm{L}+\bm{S}$, are obtained as  
\begin{align}
V_{Q\bar{Q}}(r)&=\langle {}^{2S+1}L_J| V_{Q\bar{Q}}(\bm{r}) |
                 {}^{2S+1}L_J\rangle\cr &= V(r)+ 
\left[ \frac12S(S+1)-\frac34\right] V_{SS}(r)\cr
&  + \langle \bm{L} \cdot \bm{S}\rangle  V_{LS}(r) +\langle
  \bm{\Omega}_T \rangle V_T(r).
\label{eq:PotMatrix}
\end{align}
Here the matrix element of the tensor operator is obtained to be
\begin{align}
\langle \bm{\Omega}_T\rangle&=
\frac{S(S+1)L(L+1)}{3(2L-1)(2L+3)}\cr
&-\frac{2\langle\bm{L}\cdot \bm{S}\rangle
  \left[2\langle \bm{L}\cdot \bm{S}\rangle+1\right]}{4(2L-1)(2L+3)} 
\end{align}
and $\langle \bm{L}\cdot \bm{S}\rangle$ is given by  expression
\begin{align}
 \langle\bm{L}\cdot \bm{S} \rangle = \frac12\left[J(J+1) 
  -L(L+1) -S(S+1)\right].
\end{align}

The corresponding radial part of the wave function 
for a given orbital angular momentum $L$ is a solution of the
Schr\"odinger equation  
\begin{align}
\label{eq:RadSE}
\left(-\frac{\hbar^2}{m_Q}\nabla^2+V_{Q\bar
  Q}(r)-E\right)\psi_{LL_3}(\bm{r})=0,
\end{align}
where an angular part of the wave function $\psi_{LL_3}(\bm{r})$
is represented in terms of the spherical harmonics
$Y_{LL_3}(\hat{\bm{r}})$.  

In order to solve Eq.\,\eqref{eq:RadSE} numerically, we will follow the  
Gaussian expansion method~(see review~\cite{Hiyama:2003cu}).
Firstly, we expand the state vector $|\psi_{LL_3}\rangle$ in terms of
a set of basis vectors 
$\{|\phi_{nLL_3}\rangle;\,n=1,2,\dots,n_{\mathrm{max}}\}$ as
\begin{align}
|\psi_{LL_3}\rangle=\sum_{n=1}^{n_{\rm max}}C_{n}^{(L)}|\phi_{nLL_3}\rangle.
\label{eq:Expan}
\end{align}
Secondly, we express the basis wave functions in the
spherical coordinates 
\begin{align}
\phi_{nLL_3}(\bm{r})=\phi^G_{nL}(r)Y_{LL_3}(\hat{\bm{r}}).
\label{eq:phiwf}
\end{align}
The radial part of wave function 
is expressed in terms of the Gauss functions 
\begin{align}
\phi^G_{nL}(r)=\left(\frac{2^{2L+{7}/{2}}r_{n}^{-2L-3}}{\sqrt{\pi}(2L+1)!!}
  \right)^{1/2}r^{L}e^{-(r/r_{n})^2},
\end{align}
where $r_n$ are variational parameters. 
One should note that the set of the wave functions
$\{\phi_{nLL_3}(\bm{r}); n=1,2,\dots,n_{\rm max}\}$ are 
properly normalized while they do not need to satisfy the
orthogonality condition, i.e. they consist of a non-orthogonal basis.  
To obtain high accuracy by using the expansion in Eq.~\eqref{eq:Expan}
one should optimize the set of variational parameters
$\{r_n; n=1,2,\dots,n_{\rm max}\}$. We follow an optimization
discussed in Ref.\,\cite{Hiyama:2003cu} and express the variational
parameters by using a geometric progression 
\begin{align}
r_n=r_1a^{n-1},\quad n=1,2,\dots,n_{\rm max}.
\label{eq:Gprog}
\end{align}
Thus, the number of variational parameters is reduced down to three,
i.e. $\{r_1, a, n_{\rm max}\}$ or $\{r_1, r_{\rm max}, n_{\rm
  max}\}$. 

The expansion coefficients $C_n^{(L)}$ in Eq.\,(\ref{eq:Expan})
and the eigenenergy $E$ are determined by employing the Rayleigh-Ritz
variational principle. This leads to a generalized matrix eigenvalue
problem 
\begin{align}
\sum_{n=1}^{n_{\rm max}}\left(K_{mn}^{(L)}+V_{mn}^{(L)}
-EN_{mn}^{(L)}\right)&
C_{n}^{(L)}=0,\\
 m=1,2,\dots,n_{\rm max},&\nonumber
\end{align}
where the matrix elements of the corresponding kinetic and potential
energies are obtained by
\begin{widetext}
\begin{align}
T_{mn}^{(L)}&=\langle\phi_{mLL_3}\left|\frac{\hat p^2}{m_Q}\right| 
\phi_{nLL_3}\rangle=\frac{\hbar^2(2L+1)}{m_Q(r_m^2+r_n^2)}\,
N_{mn}^{(L)},\\
V_{mn}^{(L)}&=\langle\phi_{mLL_3}|V_{Q\bar{Q}}(r)|\phi_{nLL_3}\rangle
=\frac{2^{2L+{7}/{2}}}{\sqrt{\pi}(2L+1)!!\,(r_mr_{n})^{L+3/2}}
\int_0^\infty\!\! r^{2(L+1)}\exp\left\{-\frac{r^2(r_m^2+r_n^2)}
{r_m^2r_n^2}\right\}V_{Q\bar{Q}}(r){\rm d}r,\\
N_{mn}^{(L)}&=\langle\phi_{mLL_3}|\phi_{nLL_3}\rangle=\left(\frac{2r_mr_n}
{r_m^2+r_n^2}\right)^{L+3/2}.
\end{align}
\end{widetext}

Because of Eq.\,(\ref{eq:Gprog}), the overlap matrix element 
between the nearest neighborhoods 
\begin{align}
N_{n+1,n}^{(L)}=\left(\frac{2}{1+a^2}\right)^{L+3/2}
\end{align}
is a constant that is independent of $n$. This is one of the reasons
why the expansion works very well. In a practical calculation $a>1$ and  
for the farthest neighborhoods ($|n-m|=k\gg 1$) an orthogonality 
is approximately satisfied, i.e. the corresponding overlap matrix
element becomes small, $N_{mn}^{(L)}\sim 2a^{-k-2}$.

\section{Results and discussions}
\label{sec:Results} 
In the present work, we set up three different models, depending on
how to fix the numerical values of the relevant parameters. 
The first model is merely a nonrelativistic potential model based only
on the Coulomb-like potential and the linear scalar
one~\cite{Barnes:2005pb} without any contributions from the instanton
vacuum. We will call it model without instantons (MWOI) and list the 
corresponding set of parameters in Table~\ref{tab:1}.  
\begin{table}[hbt]
\caption{Parameters corresponding to each model. 
MWOI represents the model without any instanton contributions, whereas
M-I and M-IIb contain them as explained in
Ref.~\cite{Turimov:2016adx}.}
\begin{ruledtabular}
\begin{tabular}{c|cccccc}
The & $\rho$ & $R$ &  $\Delta m_{\rm I}$ & $\alpha_s$ 
&$\kappa$ & $\sigma$\\
model &[fm]& [fm]&[GeV]&[GeV]&[GeV$^2$]&[GeV]\\
\hline
MWOI&-&-&-&0.2068&0.1746&5.0248\T\\
M-I&0.33& 1.00 &0.0676&0.3447&0.1520&0.9331\\
M-IIb&0.36& 0.89&0.1357&0.4588&0.1279&0.5650\\
\end{tabular}
\end{ruledtabular}
\label{tab:1}
\end{table} 
We do not intend to carry out any fine tuning to the experimental 
data but we concentrate on nonperturbative physics as to how the
instanton-induced potentials have an effect on the charmonium spectrum.
We also discuss physical implications of the parameters involved in
the present model, including both the instanton parameters and other
ones such as $m_Q$, $\kappa$ and $\alpha_s$. For example, while the
charm-quark mass $m_c$ is often treated as a free parameter, we will
not consider it as a free parameter. Once the instanton effects are
taken in to account, $m_c$ appears as a sum of the current quark mass
$m_c^{\mathrm{current}}=1.275$\,GeV and the dynamical contribution to
the mass arising from the instanton vacuum, $\Delta  
m_{\mathrm{I}}$, i.e. $m_Q=m_c^{\mathrm{current}} + \Delta
m_{\mathrm{I}}$ (see detailed discussions in
Refs.~\cite{Diakonov:1989un, Chernyshev:1994zm, Turimov:2016adx}). 
The remaining three parameters $\alpha_s$, $\kappa$ and $\sigma$ will
be fitted to experimentally known masses of six charmonia that are
taken from the $S$-wave ones. By doing that, we fix the parameters
appearing in the central part of the potentials, while  
those of the spin-dependent parts are automatically determined. Note
again that the charm-quark mass is not a free parameter.  
So, the spin-orbit and tensor potentials must come out  
naturally as in Eq.~\eqref{eq:VLSP} and Eq.~\eqref{eq:VLS}.  
We list the numerical values of all the parameters corresponding to
each model in Table~\ref{tab:1}, which were fixed as 
explained above.

In Table~\ref{tab:2}, we present the results of each model in
comparison with the experimental data~\cite{PDG2018} listed in the
last column. The masses of the six $S$-wave charmonia are
used as input, as noted in the second column. 
\begin{table}[hbt]
\caption{Results on the masses of the $c\bar{c}$ states, given in unit
  of MeV. The second column denotes explicitly those of the $S$-wave
  charmonia used as input to fix the parameters ($\alpha_s$, $\kappa$
  and $\sigma$) for each model. The third column lists the results
  from the original potential without the instanton-induced
  potentials, where as the fourth and fifth column show those from
  Model~I and Model~IIb, respectively. 
}
\begin{ruledtabular}
\begin{tabular}{cccccc} 
State&Input&MWOI&M-I&M-IIb&Exp.\,\cite{PDG2018} \\
\hline
$J/\psi(1^3{\rm S}_1)$&
3097&3084&3094&3096&$3096.900\pm0.006$\T\\
$\eta_c(1^1{\rm S}_0)$&
2983&3027&2998&2983&$2983.9\pm0.5$\\
$\psi(2^3{\rm S}_1)$&
3686&3635&3656&3675&$3686.097\pm0.025$\T\\
$\eta_c(2^1{\rm S}_0)$&
3640&3590&3615&3638&$3637.6\pm1.2$\\
$\psi(3^3{\rm S}_1)$&
4040&4067&4069&4071&$4039\pm1$\T\\
$\eta_c(3^1{\rm S}_0)$&
&4026&4041&4047&\\
$\psi(4^3{\rm S}_1)$&
4415&4443&4422&4398&$4421\pm 4$\T\\
$\eta_c(4^1{\rm S}_0)$&
&4405&4400&4379&\\
$\chi_{c2}(1^3{\rm P}_2)$&
&3428&3607&3740&$3556.17\pm 0.07$\T\\
$\chi_{c1}(1^3{\rm P}_1)$&
&3437&3589&3715&$3510.67\pm 0.05$\\
$\chi_{c0}(1^3{\rm P}_0)$&
&3415&3551&3673&$3414.71 \pm 0.30$\\
$h_c(1^1{\rm P}_1)$&
&3430&3599&3727&$3525.38\pm0.11$\\
$\chi_{c2}(2^3{\rm P}_2)$&
&3888&4039&4138&$3927.2\pm2.6$\T\\
$\chi_{c1}(2^3{\rm P}_1)$&
&3890&4030&4125&\\
$\chi_{c0}(2^3{\rm P}_0)$&
&3866&4006&4098& $3862_{-32-13}^{+26+40}$ \\
$h_c(2^1{\rm P}_1)$&
&3887&4039&4134&\\
$\chi_{c2}(3^3{\rm P}_2)$&
&4281&4414&4466&\T\\
$\chi_{c1}(3^3{\rm P}_1)$&
&4280&4394&4455&\\
$\chi_{c0}(3^3{\rm P}_0)$&
&4256&4375&4436&\\
$h_c(3^1{\rm P}_1)$&
&4278&4402&4463&\\
$\psi_3(1^3{\rm D}_3)$&
&3692&3830&3929&\T\\
$\psi_2(1^3{\rm D}_2)$ & 
&3718&3836&3927& $3822.2\pm 1.2$ \\
$\psi(1^3{\rm D}_1)$&
&3730&3830&3914&$3778.1\pm1.2$\\
$\eta_{c2}(1^1{\rm D}_2)$&
&3708&3837&3930&\\
$\psi_3(2^3{\rm D}_3)$&
&4104&4238&4311&\T\\
$\psi_2(2^3{\rm D}_2)$&
&4124&4242&4310&\\
$\psi(2^3{\rm D}_1)$&
&4131&4241&4303&$4191\pm5$\\
$\eta_{c2}(2^1{\rm D}_2)$&
&4116&4245&4314&\\
\end{tabular}
\end{ruledtabular}
\label{tab:2}
\end{table}
As shown from the results of the MWOI listed in the third column of
Table~\ref{tab:2}, the results on the masses of the $S$-wave charmonia
are somewhat deviated from the experimental data. If one had released
the charm-quark mass to be a mere free parameter, we would have
described the experimental data very well. For example, various
potential models adopt larger values of $m_c$ than in the present
work, since they yield better results, compared to the experimental
data. However, such a fitting procedure would obscure the physical
meaning of the quark mass $m_Q$. We want to emphasize that the
heavy-quark mass itself is a physical and dynamical quantity that 
can be also influenced by both perturbative and nonperturbative
interactions. Thus, one needs to analyze carefully the effects of the
charm-quark mass on a relevant physical system. In fact, the quarkonia
provides a system appropriate for scrutinizing the physical
implications of the charm-quark mass. In particular, the effects from
the instanton vacuum put us on the right track. Using $m_c$
with the instanton contribution added, we are able to fit better the
three parameters to the masses of the six $S$-wave charmonia. 

In the instanton liquid model of the QCD vacuum, one can estimate  
the instanton contribution to the current quark mass. In the
light-quark systems, the generation of the mass is fully dynamical, so
that almost all the constitutent quark mass originiates from the
spontaneous breaking of chiral symmetry. When it comes to the
heavy-quark systems, one needs also to consider them, though the
nonperturbative effects are not as much significant as in the case of
the light hadrons. In Table~\ref{tab:1}, the explicit values 
of $\Delta m_{\rm I}$ due to
the instanton effects are presented for each model. The strength of
the instanton effects depends on the instanton density $n\sim
(\rho/R)^4$ and the acquired mass $\Delta m_{\rm I}$ 
is proportional to it
(compare the value in M-I with that in M-IIb, listed in Table~\ref{tab:1}). 

The results of calculations presented in Table~\ref{tab:2} show that M-I describes the mass
spectrum of the charmonia quite well even up to the $D$-wave
charmonia. In particular, the mass of the $D$-wave charmonium
$\psi_2(1 {}^3D_2)$ is much improved by the instanton effects in M-I. 
On the other hand and although it gives better fit to $S$-wave states,
 M-IIb does not show any improvement in comparison
with the MWOI as related to $\psi_2(1 {}^3D_2)$. 
It indicates that the original values of the instanton
parameters $\rho$ and $R$ yield the better results, which are in fact
expected. Those values employed in M-IIb were determined when the
$1/N_c$ meson-loop corrections are involved, which is not the case of
the present work. 

Another important issue lies in the running strong coupling constant
$\alpha_s$. Since the Coulomb-like potential arises from OGE in pQCD,  
the value of $\alpha_s(\mu)$ should be taken from pQCD at a proper
scale related to the charmonia. In fact, the static Coulomb potential
was extensively studied within the framework of
pQCD, higher-order corrections being even taken into
account~\cite{Peter:1996ig,Peter:1997me, Schroder:1998vy, 
Smirnov:2009fh, Anzai:2009tm}. However, a fitted value of $\alpha_s$
was used in many models because it is simpler and phenomenologically
more favorable than that from pQCD. At the one-loop level it is given
by  expression  
\begin{align}
\alpha_s(\mu)=\frac{4\pi}{\beta_0}\frac{1}{\ln\left(\mu^2/
  \Lambda_{\mathrm{QCD}}^2\right)},  
\label{alphas}
\end{align}
where the $\beta$ function at the one-loop level is equal to be
$\beta_0=({11}N_c-2N_f)/3$. The dimensionful parameter of QCD is given 
as $\Lambda_{\mathrm{QCD}}=0.217$\,GeV~\cite{PDG2018} and  
$\mu$ stands for a specific scale at which the value of $\alpha_s$ is
evaluated. It is usual to take $\mu\approx m_c$ for the
charmonia. 

The value of $\alpha_s$ at the one-loop level can be easily computed
by using Eq.\,\eqref{alphas}, once a proper scale $\mu$ is given. 
Considering the fact that MWOI, M-I and M-IIb have different
charm-quark masses, we can calculate the value of
$\alpha_s$ corresponding to a specific model by using the charm-quark
mass of the model as its intrinsic scale. The results for M-I and
M-IIb are obtained respectively as
$\alpha_s(\mu=1.343\,\mathrm{GeV})=0.4137$ and
$\alpha_s(\mu=1.411\,\mathrm{GeV})=0.4029$, which 
are slightly larger or smaller 
than the corresponding fitted values listed in
Table~\ref{tab:1}, respectively.  On the other hand, the value of $\alpha_s$ for
MWOI turns out to be $\alpha_s(\mu=1.275\,\mathrm{GeV})=0.4258$, which 
is approximately two times larger than the fitted value $0.2068$. This
comparison already demonstrates that in the presence of the 
instanton-induced interaction one is allowed to use a more
\textit{physical} strong coupling constant for the Coulomb-like
potential, based firmly on perturbative QCD. 

The origin of the confining scalar potential is not much known
theoretically and the string tension $\kappa$ is considered as a
phenomenological parameter, though it can be related to Regge
trajectories within some models~\cite{Bali:2000gf}. The parameter
$\kappa$ was often determined by the charmonium spectrum together with  
the \textit{effective} strong coupling constant~\cite{Eichten:1978tg,
  Quigg:1979vr, Bali:2000gf} and the numerical value of $\kappa$ is
known approximately to be $\kappa\approx 0.18\,\mathrm{GeV}^2$ (see a 
review~\cite{Bali:2000gf}). In the present work, we fix it to be
$0.175\,\mathrm{GeV}^2$ for the MWOI, which is very close to the
above-given value. However, once we introduce the instanton-induced
potential, we have to use smaller values of $\kappa$, because the
central part of instanton-induced potential 
has almost a linearly rising behavior and then is
saturated as the inter-distance of the quarks
increases~\cite{Diakonov:1989un, Turimov:2016adx}. It means that 
the instanton effects reduce the strength of the linear scalar
potential. In other words, the instanton-induced interactions 
contribute partially to the confining potential in a general form
\begin{align}
V_{\rm conf}(r) =\mbox{const.}+\kappa r + \mbox{possible nonlinear
  terms.} 
\label{Vconf}
\end{align}
For example,  
the leading order contribution from instantons in 
Eq.\,(\ref{eq:LO_pot}) to the total $Q\bar{Q}$ potential at small 
distances can be treated as a nonlinear corrections to the 
linearly raising potential. In this context, the instanton 
induced potential at large distances 
$(4\pi\rho^{3}/{N_cR^4})\mathcal{I}_0$  
has a meaning 
of partial contribution to the constant part of confining potential 
in Eq.\,(\ref{Vconf}).

Finally, we discuss the smearing parameter $\sigma$ that was
introduced to avoid the singular behavior of the point-like spin-spin
interactions. Its value gets smaller if the
instanton-induced potentials are included. Note that the point-like
interaction is actually an artifact of an ${\cal O}(v^2_q/c^2)$
expansion of the $T$-matrix~\cite{Barnes:1982eg}. Introducing the 
instanton-induced potential~\footnote{Note that the spin-dependent
  parts of the instanton-induced potentials are regular.}, one can
partially solve the divergence problem that arises from the singular
spin-spin interactions.

\section{Summary and outlook \label{sect:Summary}}
In the present work, we aimed at investigating the instanton effects
on the charmonium mass spectrum, based on a nonrelativistic
potential model. Though we fixed the relevant parameters by using the
masses of the six $S$-wave charmonia, we did not intend to carry out
the fine-tuning of the parameters. The results showed that the
instanton-induced potentials improved the mass spectrum with the
original values of the average size of the instanton and the average
inter-distance between instantons. We discussed also the physical
implications of the parameters such as the charm-quark mass, the
running strong coupling constant, the string tension, and the smearing
factor. The instanton-induced potentials enable one to understand more
clearly these physical parameters. 

Though we considered certain nonperturbative contributions to the mass
spectrum of the charmonia from the instanton vacuum, we still need to
take into account yet additional effects from the
instantons. Recently, it was shown that 
the instanton effects or the screening effects
in the Coulomb-like potential of one-gluon
exchange will appear due to gluon propagation in instanton 
media~\cite{Musakhanov:2017erp}. Since the 
gluon is screened by the instanton effects and as a result it acquires
an effective mass, we have in addition a Yukawa-type potential between 
the heavy quark and heavy anti-quark. One needs to examine how this
screened potential may influence the charmonium masses.  

Another interesting issue is related to strong decays of excited
charmonia, which involve the pions. Since the pion is the
pseudo-Goldstone boson that appears from the spontaneous breakdown of
chiral symmetry, the instanton effects will come into critical play in
describing these decay processes. Related works are under way. 

\section*{Acknowledgments}
We are grateful to P. Gubler, A. Hosaka, T. Maruyama, M. Oka and  
M. Musakhanov for useful discussions.  H.-Ch.K wants to express his
gratitude to the members of the Advanced Science Research Center
(ASRC) at Japan Atomic Energy Agency (JAEA) for the hospitality, where
part of the present work was done.  This work is supported by the
Basic Science Research Program through the National Research
Foundation (NRF) of Korea funded by the Korean 
government (Ministry of Education, Science and Technology, MEST),
Grant Numbers~2016R1D1A1B03935053 (UY) and
NRF2018R1A2B2001752~(HChK). The work was also partly supported by
RIKEN iTHES Project.

\end{document}